\documentclass[11pt]{article}
\usepackage{epsfig}
\usepackage{amssymb,amsmath,amsthm,url,verbatim}
\usepackage{multirow}
\usepackage{booktabs}
\usepackage{setspace}
\usepackage{dsfont}

\usepackage{times,amsfonts,eucal}
\usepackage{algorithm}
\usepackage{algorithmic}
\usepackage{graphicx,ifthen}
\usepackage{wrapfig}
\usepackage{latexsym}
\usepackage{color}
\usepackage{mathrsfs}
\usepackage{bm}
\usepackage{bbm}
\usepackage{srctex}

\usepackage{bm}
\newtheorem{theorem}{Theorem}[section]
\newtheorem{lemma}{Lemma}[section]
\newtheorem{pro}{Proposition}[section]
\newtheorem{cor}{Corollary}[section]

\newtheorem{assumption}{Assumption}[section]
\numberwithin{equation}{section}
\numberwithin{theorem}{section}
\numberwithin{lemma}{section}
\numberwithin{pro}{section}
\numberwithin{cor}{section}
\numberwithin{definition}{section}
\numberwithin{cons}{section}
\numberwithin{rem}{section}
\numberwithin{exa}{section}
\numberwithin{table}{section}
\numberwithin{figure}{section}

\newcommand{\intt}{\int\hspace{-.2cm}\int}

\def\beq{\begin{equation}}
\def\eeq{\end{equation}}
\def\bals{\begin{align*}}
\def\eals{\end{align*}}

\def\bal{\begin{align}}
\def\eal{\end{align}}

\allowdisplaybreaks

\begin{document}

\title{Detecting and dating structural breaks in functional data\\[.2cm] without dimension reduction\footnote{This research was partially supported by NSF grants DMS 1209226, DMS 1305858 and DMS 1407530. The authors would like to thank the editor, David Dunson, the associate editor and three referees for constructive criticism that has led to a substantial improvement in the quality of this paper.
}}

\author{
Alexander Aue\footnote{Department of Statistics, University of California, Davis, CA 95616, USA, email: \tt{[aaue,osonmez]@ucdavis.edu}}
\and Gregory Rice\footnote{Department of Statistics and Actuarial Science, University of Waterloo, Waterloo, ON, Canada, email: \tt{grice@uwaterloo.ca}}
\and Ozan S\"onmez$^\dagger$
}
\date{\today}
\maketitle

\begin{abstract}
\setlength{\baselineskip}{1.72em}
Methodology is proposed to uncover structural breaks in functional data that is ``fully functional" in the sense that it does not rely on dimension reduction techniques. A thorough asymptotic theory is developed for a fully functional break detection procedure as well as for a break date estimator, assuming a fixed break size and a shrinking break size. The latter result is utilized to derive confidence intervals for the unknown break date. The main results highlight that the fully functional procedures perform best under conditions when analogous fPCA based estimators are at their worst, namely when the feature of interest is orthogonal to the leading principal components of the data. The theoretical findings are confirmed by means of a Monte Carlo simulation study in finite samples. An application to annual temperature curves illustrates the practical relevance of the proposed procedures.
\medskip


\end{abstract}

\setlength{\baselineskip}{1.708em}

\section{Introduction}
\label{sec:intro}

This paper considers the problem of detecting and dating structural breaks in functional time series data, and hence lies at the intersection of functional data analysis (FDA) and structural breaks analysis for dependent observations. FDA has witnessed an upsurge in research contributions in the past decade. These are documented, for example, in the comprehensive books by Ramsay and Silverman (2005) and Ferraty and Vieu (2010). Research concerned with structural breaks has a longstanding tradition in both the statistics and econometrics communities. Two recent reviews by Aue and Horv\'ath (2013) and Horv\'ath and Rice (2014) highlight newer developments, the first with a particular focus on time series.

Early work in functional structural break analysis dealt primarily with random samples of independent curves, the question of interest being whether all curves have a common mean function or whether there are two or more segments of the data that are homogeneous within but heterogeneous without. Berkes et al.\ (2009) developed statistical methodology to test the null hypothesis of no structural break against the alternative of a (single) break in the mean function assuming that the error terms are independent and identically distributed curves. Aue et al.\ (2009) quantified the large-sample behavior of a break date estimator under a similar set of assumptions. The work in these two papers was generalized by Aston and Kirch (2012a,\,b) and Torgovitski (2016) to include functional time series exhibiting weak dependence into the modeling framework. In Zhang et al.\ (2011), a structural break detection procedure for serially correlated functional time series data is proposed that is based on the self-normalization approach of Shao and Zhang (2010). Structual break detection in the context of functional linear models is considered in Aue et al.\ (2014) and for spatially distributed functional data in Gromenko et al.\ (2016). Smooth deviations from stationarity of functional time series in the frequency domain were studied in Aue and van Delft (2017+). 

Most of the procedures in FDA, such as those presented in the above cited papers, are based on dimension reduction techniques, primarily using the widely popular functional principal components analysis (fPCA), by which the functional variation in the data is projected onto the directions of a small number of principal curves, and multivariate techniques are then applied to the resulting sequence of score vectors. This is also the case in functional structural break detection, in which after an initial fPCA step multivariate structural break theory is utilized. Despite the fact that functional data are, at least in principle, infinite dimensional, the state of the art in FDA remains to start the analysis with an initial dimension reduction procedure.

Dimension reduction approaches, however, automatically incur a loss of information, namely all information about the functional data that is orthogonal to the basis onto which it is projected. This weakness is easily illustrated in the context of detecting and dating structural breaks in the mean function: if the function representing the mean break is orthogonal to the basis used for dimension reduction, there cannot be a consistent test or estimator for the break date in that basis. This point will be further illustrated by theoretical arguments and in comprehensive numerical studies in Section \ref{sec:sim}, where other more subtle differences between the competing methods will be highlighted.

The main purpose of this paper is then to develop methodology for detecting and dating structural breaks in functional data without the application of dimension reduction techniques. Here, fully functional test statistics and break date estimators are studied, and their asymptotic theory is developed under the assumption that the model errors satisfy a general weak dependence condition. This theory illuminates a number of potential advantages of the fully functional procedures. For example, it is shown that when the direction of the break is orthogonal to the leading principal components of the data, the estimation of the mean break is asymptotically {\it improved} when using the fully functional estimator compared to mean breaks of the same size that are contained in the leading principal components. This contrasts with fPCA based techniques in which such mean breaks are more difficult, if not impossible, to detect, even given arbitrarily large sample sizes. In addition, the assumptions required for the fully functional theory are weaker than the ones used in Aue et al.\ (2009) and Aston and Kirch (2012a,\,b), as convergence of the eigenvalues and eigenfunctions of the empirical covariance operator to the eigenvalues of the population covariance operator do not have to be accounted for. These assumptions are typically formulated as finiteness of fourth moment conditions. The relaxation obtained here may be particularly useful for applications to intra-day financial data such as the one-minute log-returns on Microsoft stock discussed in the online supplement Aue et al.\ (2017+) accompanying this article.

The application presented in Section \ref{sec:app} is concerned with annual temperature curves recorded across different measuring stations in Australia. Structural breaks in these temperature curves are detected with both fPCA and fully functional methods. The sample covariance operator associated with the data has eigenvalues that decay remarkably slowly. A somewhat peculiar feature of fPCA methods in this setting, studied as part of the simulation experiment, is a loss of accuracy in break dating even when the break function loads almost exclusively on the first component. A similar effect is found in the data, where fPCA-based break dates can occur outside of the confidence intervals provided by the fully functional procedure.

Most closely related to the present work are Fremdt et al.\ (2014), who considered structural break detection using fPCA under an increasing number of projections. Horv\'ath et al.\ (2014) developed a functional analog of the KPSS test statistic for the purpose of stationarity testing that does not rely on dimension reduction. Sharipov et al.\ (2016) considered a bootstrap procedure for measuring the significance of the norms of functional CUSUM processes with applications to testing for a structural break in the means of functional observations and in the distribution function of scalar time series observations under a mixing assumption, generalizing the result for the independent, identically distributed case put forward in Tsudaka and Nishiyama (2014). Bucchia and Wendler (2016+) studied general bootstrap procedures for structural break analysis in Hilbert space-valued random fields. 



The remainder of the paper is organized as follows. Testing procedures and a break date estimator are introduced in Section \ref{sec:main}, along with the main asymptotic results of the paper. The asymptotic properties developed in this section are accompanied by implementation details given in Section \ref{sec:imp} and results from a comprehensive simulation study in Section \ref{sec:sim}. The application to temperature curves is given in Section \ref{sec:app}, and Section \ref{sec:conclusions} concludes. Proofs of the main results as well as additional empirical illustrations of the proposed methodology are provided in the online supplement Aue et al.\ (2017+), henceforth referred to simply as the online supplement. In addition, an R package, {\tt{fChange}}, has been developed to supplement this article and is available on the Comprehensive R Archive Network. The package contains implementations of all of the testing and estimation procedures introduced below, see S\"onmez et al. (2017).

\section{Main results}
\label{sec:main}

In this paper, a functional data model allowing for a mean function break is considered. It is assumed that the observations $X_1,\ldots,X_n$ are generated from the model
\begin{align}\label{model-1}
X_i= \mu+ \delta\mathds{1}\{i > k^*\} + \varepsilon_i,
\qquad i \in\mathbb{Z},
\end{align}
where $k^*=\lfloor \theta n\rfloor$, with $\theta\in(0,1)$, labels the unknown time of the mean break parameterized in terms of the sample size $n$, $\mu$ is the baseline mean function that is distorted by the addition of $\delta$ after the break time $k^*$, $\mathds{1}_A$ denotes the indicator function of the set $A$ and $\mathbb{Z}$ the set of integers. Each $X_i$ is a real-valued function defined without loss of generality on the unit interval $[0,1]$. The argument $t\in[0,1]$ will be used to refer to a particular value $X_i(t)$ of the function $X_i$. Correspondingly, the quantities $\mu$, $\delta$ and $\varepsilon_i$ on the right-hand side of \eqref{model-1} are functions on $[0,1]$ as well. Interest is first in testing the structural break hypotheses
\[
H_0\colon \delta=0
\qquad\mbox{versus}\qquad
H_A\colon \delta \ne 0,
\]
and then, in the event that $H_A$ is thought to hold, estimating the break date $k^*$. Throughout the following assumptions are made, roughly entailing that the innovations $(\varepsilon_i\colon i\in\mathbb{Z})$ are weakly dependent, stationary functional time series. Below, let $\|\cdot\|$ and $\langle \cdot, \cdot \rangle$ denote the canonical norm and inner product in $L^2[0,1]$.

\begin{assumption}\label{edep}
The innovations $(\varepsilon_i\colon i\in\mathbb{Z})$ satisfy

(a) there is a measurable function $g\colon S^\infty\to L^2[0,1]$, where $S$ is a measurable space and independent, identically distributed (iid) innovations $(\epsilon_i\colon i\in\mathbb{Z})$ taking values in $S$ such that $\varepsilon_i=g(\epsilon_i,\epsilon_{i-1},\ldots)$ for $i\in\mathbb{Z}$;

(b) there are $\ell$-dependent sequences $(\varepsilon_{i,\ell}\colon i\in\mathbb{Z})$ such that, for some $p>2$,
\[
\sum_{\ell=0}^\infty\big(\mathbb{E}[\|\varepsilon_i-\varepsilon_{i,\ell}\|^p]\big)^{1/p}<\infty,
\]
where $\varepsilon_{i,\ell}=g(\epsilon_i,\ldots,\epsilon_{i-\ell+1},\epsilon^*_{i,\ell,i-\ell},\epsilon^*_{i,\ell,i-\ell-1},\ldots)$ with $\epsilon^*_{i,\ell,j}$ being independent copies of $\epsilon_{i,0}$ independent of $(\epsilon_i\colon i\in\mathbb{Z})$.
\end{assumption}
Processes satisfying Assumption \ref{edep} were termed $L^p$-$m$-approximable by H\"ormann and Kokosz\-ka (2010), and cover most stationary functional time series models of interest, including functional AR and ARMA (see Aue et al., 2015; and Bosq, 2000) and functional GARCH processes (see Aue et al., 2017). It is assumed that the underlying error innovations $(\epsilon_i\colon i\in\mathbb{Z})$ are elements of an arbitrary measurable space $S$. However, in many examples $S$ is itself a function space, and the evaluation of $ g(\epsilon_{i},\epsilon_{i-1},...)$ is a functional of $(\epsilon_j\colon j\leq i)$.

The proposed methodology is based on the (scaled) functional cumulative sum (CUSUM) statistic
\begin{align}\label{s0def}
S_{n,k}^0= \frac{1}{\sqrt{n}} \bigg( \sum_{i=1}^k X_i - \frac{k}{n}\sum_{i=1}^n X_i \bigg).
\end{align}
The superscript 0 indicates the tied-down nature of the CUSUM statistic, since $S_{n,0}^0=S_{n,n}^0=0$ (interpreting an empty sum as zero). Noting that $\|S_{n,k}^0\|$ as a function of $k$ tends to be large at the true break date motivates the use of the max-type structural break detector
\[
T_n = \max_{1 \le k \le n} \| S_{n,k}^0\|^2
\]
to test $H_0$ versus $H_A$. Furthermore, the break date estimator for $k^*$  may be taken as
\begin{align}\label{estdef}
\hat{k}^*_n = \min\Big\{k\colon \|S_{n,k}^0\|=\max_{1\leq k^\prime\leq n}\|S_{n,k^\prime}^0\|\Big\}.
\end{align}
The main results of this paper concern the large-sample behavior and empirical properties of the test statistic $T_n$ and the estimator $k_n^*$.

\subsection{Asymptotic properties of structural break detector}

Under $H_0$, the limiting behavior of $S_{n,k}^0$ evidently depends on that of the partial sum process of the error terms $(\varepsilon_i\colon i\in\mathbb{Z})$. As this sequence may be weakly serially correlated under Assumption \ref{edep}, the asymptotics of the partial sum process necessarily involve the long-run covariance kernel
\begin{align}\label{Ddef}
C_\varepsilon(t,t^\prime) = \sum_{\ell=-\infty}^\infty \mathrm{Cov}(\varepsilon_0(t), \varepsilon_\ell(t^\prime))
\end{align}
of the error sequence $(\varepsilon_i\colon i\in\mathbb{Z})$. Note that $C_\varepsilon$ constitutes the limiting covariance kernel of $\sqrt{n}$ times the centered sample mean under $H_0$. It is a well-defined element of $L^2[0,1]^2$ under Assumption \ref{edep}. This kernel was considered initially in H\"ormann and Kokosz\-ka (2010). It was also studied in Panaretos and Tavakoli (2012) in the context of spectral analysis of functional time series, and in Horv\'ath et al.\ (2013) in an application to the functional two sample problem. In addition, $C_\varepsilon$ may be used to define a positive definite and symmetric Hilbert--Schmidt integral operator on $L^2[0,1]$, $c_\varepsilon$, given by
\[
c_\varepsilon(f)(t) = \int C_\varepsilon(t,s)f(s)ds,
\]
which further defines a non-increasing sequence of nonnegative eigenvalues $(\lambda_\ell\colon\ell\in\mathbb{N})$ and a corresponding orthonormal basis of eigenfunctions $(\phi_\ell\colon\ell\in\mathbb{N})$ satisfying
\begin{equation}\label{eigen-eq}
c_\varepsilon(\phi_\ell)(t) = \lambda_\ell\phi_\ell(t),
\qquad \ell\in\mathbb{N}.
\end{equation}
The eigenvalues of $c_\varepsilon$ determine the limiting distribution of $T_n$ as detailed in the following theorem.

\begin{theorem}\label{inf-main} Under Model \ref{model-1}, Assumption \ref{edep} and $H_0$,
\begin{align}\label{tn-1}
T_n \stackrel{\cal D}{\to}\sup_{0\le x \le 1}  \sum_{\ell=1}^\infty \lambda_\ell B_\ell^2(x)
\qquad (n\to\infty),
\end{align}
where $(B_\ell\colon\ell\in\mathbb{N})$ are independent and identically distributed standard Brownian bridges defined on $[0,1]$.
\end{theorem}

Theorem \ref{inf-main} points to an asymptotically validated test of $H_0$, namely to reject if the test statistic $T_n$ exceeds the corresponding quantile of the distribution on the right hand side of \eqref{tn-1}. As the limiting distributions depends, in a rather complicated way, on the unknown eigenvalues $(\lambda_\ell\colon\ell\in\mathbb{N})$ and standard Brownian bridges, Monte Carlo simulation can be used to approximate this distribution using estimated eigenvalues. Implementation details are provided in Section \ref{sec:imp} below. Theorem 2.1 was also obtained in Sharipov et al. (2016) under a strong mixing condition that is analogous to Assumption \ref{edep}. These authors further developed a block bootstrap methodology to approximate the limiting distribution.

A common assumption made in order for analogous break point detection procedures based on fPCA to be consistent, as studied for example in Berkes et al.\ (2009) and Aston and Kirch (2012a), is that $\delta$ is not orthogonal to the principal component basis used to perform the dimension reduction step. When using the detector $T_n$ no such assumption is needed.

\begin{theorem}\label{th-alt}
Under Model \ref{model-1}, Assumption \ref{edep} and $H_A$,   $T_n \stackrel{P}{\to}\infty,$ as $n\to \infty$.
\end{theorem}

The proofs of Theorems \ref{inf-main} and \ref{th-alt} are in the online supplement.

\subsection{Asymptotic properties of the break date estimator}

Further advantages of the fully functional approach become apparent when studying the asymptotic properties of the break date estimator $\hat k_n^*$, which are established below. Two cases are studied: the fixed break situation for which the break size is independent of the sample size, and the shrinking break situation for which the break size converges to zero at a specified rate. In the fixed break case, the following holds.

\begin{theorem}\label{th-1}
If model \eqref{model-1} holds with $0\not=\delta\in L^2[0,1]$, and if Assumption \ref{edep} is satisfied, then
\begin{align}\label{th-1-eq1}
\hat{k}^*_n - k^*
\stackrel{{\mathcal D}}{\to}
\min\Big\{k\colon P(k)= \sup_{k^\prime\in\mathbb{Z}} P(k^\prime)\Big\}
\qquad(n\to\infty),
\end{align}
where
\begin{align}\label{th-1-eq2}
   P(k)=\left\{
     \begin{array}{r@{\qquad}l}
      \displaystyle (1-\theta)\|\delta\|^2k+ \langle\delta,S_{\varepsilon,k}\rangle, &k < 0, \\[.2cm]
     \displaystyle -\theta\|\delta\|^2k+ \langle\delta,S_{\varepsilon,k}\rangle, &k \geq 0,
     \end{array}
   \right.
\end{align}
with
\[
S_{\varepsilon,k}= \sum_{i=1}^k \varepsilon_i + \sum_{i=-k}^{-1} \varepsilon_i.
\]
\end{theorem}

As one can see in \eqref{th-1-eq1}, the limit distribution of $\hat{k}_n^*$ in the case of a fixed break size depends on the unknown underlying distribution of the error process. This encourages the consideration of a break  $\delta_n$ that shrinks as a function of the sample size, in which case the limit distribution is the supremum of a two-sided Brownian motion with triangular drift depending on a small set of nuisance parameters, but not otherwise on the distribution of the error sequence $(\varepsilon_i\colon i\in\mathbb{Z})$.

\begin{theorem}\label{th-2}
If model \eqref{model-1} holds with $0\not=\delta=\delta_n\in L^2[0,1]$ such that $\|\delta_n\| \to 0$ but $n\|\delta_n\|^2 \to \infty$ and if Assumption \ref{edep} is satisfied, then
\[
\|\delta_n\|^2 \big(\hat{k}_n^* - k^*\big)
\stackrel{{\mathcal D}}{\to}
\inf\Big\{ x \colon Q(x) = \sup_{x^\prime\in\mathbb{R}} Q(x^\prime)\Big\}
\qquad(n\to\infty),
\]
where $\mathbb{R}$ denotes the real numbers and
\begin{align}\label{th-2-eq2}
   Q(x)=\left\{
     \begin{array}{r@{\qquad}l}
     \displaystyle (1-\theta)x+ \sigma W(x),& x < 0, \\[.2cm]
     \displaystyle -\theta x+ \sigma W(x), & x \geq 0,
     \end{array}
   \right.
\end{align}
with $(W(x)\colon x\in\mathbb{R})$ a two-sided Brownian motion, and
$$
\sigma^2 = \lim_{n \to \infty} \intt C_\varepsilon(t,t^\prime) \frac{\delta_n(t) \delta_n(t^\prime)}{\|\delta_n\|^2} dtdt^\prime,
$$
where $C_\varepsilon(t,t^\prime)$ is the long-run covariance kernel of $(\varepsilon_i\colon i\in\mathbb{Z})$ given in \eqref{Ddef}.
\end{theorem}

An interesting consequence of Theorems \ref{th-1} and \ref{th-2} is that mean changes $\delta$ that are orthogonal to the primary modes of variation in the data are asymptotically {\it easier} to detect and estimate. For example, if, under the conditions of Theorem \ref{th-1}, $\delta$ is orthogonal to the error functions, then the stochastic term in the limit distribution vanishes. Moreover, if the functions $\delta_n$ in Theorem \ref{th-2} tend to align with eigenfunctions corresponding to smaller and smaller eigenvalues of the integral operator with kernel $C_\varepsilon$, then $\sigma^2$ tends to zero in the definition of $Q(x)$. The proofs of Theorems \ref{th-1} and \ref{th-2} are given in the online supplement.

Theorem \ref{th-2} suggests a confidence interval for $k^*$.
\begin{cor}
\label{cor:ci}
Let\/ $\Xi=\inf\{ x \colon Q(x) = \sup_{x^\prime\in\mathbb{R}} Q(x^\prime)\}$. Then, under the conditions of Theorem \ref{th-2} and for $\alpha \in (0,1)$, the random interval
\begin{align}\label{ci-f}
\left( \hat{k}^*_n - \frac{\Xi_{1-\alpha/2}}{\|\delta_n\|^{2}},\hat{k}^*_n - \frac{\Xi_{\alpha/2}}{\|\delta_n\|^{2}}\right)
\end{align}
is an asymptotic $1-\alpha$ sized confidence interval for $k^*$, where $\Xi_{q}$ is the $q$th quantile of\/ $\Xi$.
\end{cor}
The main crux here is that $\delta_n$ is unknown and the distribution of $\Xi$ depends on the unknown break fraction $\theta$ and the limiting variance parameter $\sigma^2$. Consistent estimation techniques for these parameters are discussed in Section \ref{sec:imp} below. This confidence interval tends to be conservative in practice due to the fact that it is derived under the assumption of a shrinking break. A thorough empirical study of the break date estimator and the corresponding confidence interval is provided in Section \ref{sec:sim}.

The last result of this section concerns the large-sample behavior of $\hat k_n^*$ if no break is present in the data, that is, if $\delta=0$ in \eqref{model-1}.

\begin{theorem}\label{th-3}
If model \eqref{model-1} holds with $\delta=0$, so that $X_i=\mu_i+\varepsilon_i$ for all $i=1,\ldots,n$, and if Assumption \ref{edep} is satisfied, then
\begin{align*}
\frac{\hat{k}^*_n}{n}
\stackrel{{\mathcal D}}{\to}
\arg\max_{0\le x \le 1} \| \Gamma^0(x,\cdot)\|
\qquad(n\to\infty),
\end{align*}
where $\Gamma^0$ is a bivariate Gaussian process with mean zero and covariance function
$
\mathbb{E}[\Gamma^0(x,t)\Gamma^0(x^\prime,t^\prime)]
=(\min\{x,x^\prime\}-xx^\prime) C_\varepsilon(t,t^\prime).
$
\end{theorem}

The proof of Theorem \ref{th-3} is provided in the online supplement. Observe that the limiting distribution in Theorem \ref{th-3} is non-pivotal, but it can be approximated via Monte Carlo simulations using an estimator of $C_\varepsilon$. To see this note that, because of the Karhunen--Lo\'eve representation, $\Gamma^0(x,t)$ can be written in the form $\sum_{\ell=1}^\infty\sqrt{\lambda_{\ell}}\phi_{\ell}(t)B_\ell(x)$, where $(\lambda_{\ell}\colon\ell\in\mathbb{N})$ and $(\phi_{\ell}\colon\ell\in\mathbb{N})$ are the eigenvalues and eigenfunctions of $C_\varepsilon$ and $(B_\ell\colon \ell\in\mathbb{N})$ are independent standard Brownian bridges. Computing the norm as required for the limit in Theorem \ref{th-3} yields that
\[
\arg\max_{x\in[0,1]}\| \Gamma^0(x,\cdot)\|
\stackrel{{\cal D}}{=}
\arg\max_{x\in[0,1]}\bigg(\sum_{\ell=1}^\infty\lambda_{\ell} B_\ell^2(x)\bigg)^{1/2}.
\]
Truncation of the sum under the square-root on the right-hand side gives then approximations to the theoretical limit. For practical purposes population eigenvalues have to be estimated from the data. This can be done following the steps described in Section \ref{subsec:imp:critical}.

\subsection{Two fPCA based approaches}
\label{sec:fpca_methods}

In the remainder of this section, the fully functional results put forward here are compared to their fPCA counterparts in Berkes et al.\ (2009), Aue et al.\ (2009), Aston and Kirch (2012a,\,b), and Torgovitski (2016). Berkes et al.\ (2009) and Torgovitski (2016) dealt with detection procedures and Aue et al.\ (2009) with break dating procedures, while Aston and Kirch (2012a, b) presented both. A short summary of the different approaches follows.

The works of Berkes et al.\ (2009), Aue et al.\ (2009) and Aston and Kirch (2012a,\,b) utilized the eigenvalues, say $\hat\tau_1,\ldots,\hat\tau_n$, and eigenfunctions, say $\hat\psi_1,\ldots,\hat\psi_n$, of the sample covariance operator $\hat K$ of the observations whose kernel is given by $\hat K(t,t^\prime)=n^{-1}\sum_{i=1}^n[X_i(t)-\bar X_n(t)][X_i(t^\prime)-\bar X_n(t^\prime)]$. In the presence of a mean break as in \eqref{model-1}, $\hat K(t,t^\prime)$ converges as the sample size tends to infinity to the covariance kernel
\[
K(t,t^\prime)=K_0(t,t^\prime)+\theta(1-\theta)\delta(t)\delta(t^\prime),
\]
where $K_0(t,t^\prime)=\mathbb{E}[\varepsilon_1(t)\varepsilon_1(t^\prime)]$ is the covariance kernel of the innovations $(\varepsilon_i\colon i\in\mathbb{Z})$. In particular, the eigenvalues and eigenfunctions of $\hat K(t,t^\prime)$ converge to those of $K(t,t^\prime)$ under appropriate assumptions that include the finiteness of the fourth moment $\mathbb{E}[\|\varepsilon_1\|^4]$. Choosing a suitable dimension $d\in\{1,\ldots,n\}$ allows one to define an fPCA detector based on the maximally selected quadratic form statistic
\begin{equation}\label{eq:fPCA_detector}
\tilde R_{n}
=\max_{1\leq k\leq n}\tilde{R}_{n,k}
=\max_{1\leq k\leq n}\frac 1n\tilde S_{n,k}^T\hat\Sigma_n^{-1}\tilde S_{n,k},
\end{equation}
and the break point estimator
\begin{equation}\label{eq:k-tilde}
\tilde k_n^*=\min\Big\{k\colon \tilde R_{n,k}=\max_{1\leq k^\prime\leq n}R_{n,k^\prime}\Big\},
\end{equation}
where $\tilde S_{n,k}=\sum_{i=1}^k\hat\xi_i-kn^{-1}\sum_{i=1}^n\hat\xi_i$ and $\hat\xi_i=(\hat\xi_{i,1},\ldots,\hat\xi_{i,d})^T$ with fPCA scores $\hat\xi_{i,\ell}=\langle X_i-\bar X_n,\hat\psi_\ell\rangle$, and $\hat\Sigma_n=\mathrm{diag}(\hat\tau_1,\ldots,\hat\tau_d)$. For the independent case, the counterparts of Theorems \ref{inf-main} and \ref{th-alt} were established in Berkes et al.\ (2009) for a Cram\'er--von Mises test statistics, and those of Theorems \ref{th-1} and \ref{th-2} in Aue et al.\ (2009). Aston and Kirch (2012a) considered versions of the test in \eqref{eq:fPCA_detector} and showed the consistency of $\tilde k_n^*$ in the time series case. The performance of $R_n$ and $\tilde k_n^*$ depends crucially on the selection of $d$ and the complexity of the break function $\delta$.
To briefly illustrate this point, suppose that $K_0(t,t')= \alpha b(t)b(t')$ for some orthonormal function $b$ that is orthogonal to $\delta$. It then follows from elementary calculations that $\hat{\psi}_1$ will be asymptotically orthogonal to $\delta$ if and only if $\alpha> \theta(1-\theta) \|\delta\|^2$, and hence under this latter condition one cannot have a consistent fPCA based test or break date estimator if $d=1$.
The use of the fully functional approach to dating break points is therefore especially advantageous in the interesting case of breaks that are sizable but not obvious in the sense that their influence does not show up in the directions of the leading principal components of the data.

This fact was noticed by Torgovitski (2016), who extended the detection procedures in two ways. First, instead of using the spectral decomposition of the covariance operator $K$, his procedures are based on the long-run covariance operator $C_\varepsilon$ and its eigenvalues $\lambda_1,\ldots,\lambda_n$ and eigenfunctions $\phi_1,\ldots,\phi_n$. Second, an alignment is introduced that shifts the detection procedure into the subspace of the potential break, the idea being to significantly improve power, while not majorly compromising the level. The alignment is obtained by modifying the first sample eigenfunction $\hat\phi_1$ using
\begin{equation}
\label{eq:aligned}
\tilde\phi_1^\prime
=\frac{\hat\phi_1}{n^\gamma}+\frac{\hat s\tilde S_{n,\tilde k_n^*}}{\sqrt{n}},
\end{equation}
where $\gamma\in(0,1/2)$ is a tuning parameter and $\hat s=\mathrm{sign}\langle\hat\phi_1,\tilde S_{n,\tilde k_n^*}\rangle$. Torgovitski (2016) then proposed to replace $\hat\phi_1$ with $\hat\phi_1^\prime=\tilde\phi_1^\prime/\|\tilde\phi_1^\prime\|$ in the definition of \eqref{eq:fPCA_detector}, but did not introduce the corresponding break dating procedure.

\section{Implementation details}
\label{sec:imp}

\subsection{Estimation of long-run covariance operator}
\label{subsec:imp:long-run}

The implementation of the detection procedure and confidence intervals based on the break point estimator requires the estimation of the covariance operator $C_\varepsilon$. Due to its definition as a bi-infinite sum of the lagged autocovariances of the functional time series $(\varepsilon_i\colon i\in\mathbb{Z})$, the following lag-window estimator is used. Let
\begin{align}\label{est-1}
\hat{C}_\varepsilon(t,t^\prime)=\sum_{\ell=-\infty}^{\infty}w_{\tau} \left( \frac{\ell}{h} \right) \hat{\gamma}_\ell(t,t^\prime),
\end{align}
where the components of this estimator are defined as follows: $h$ is a bandwidth parameter satisfying $h=h(n)$, and $1/h(n) + h(n)/n^{1/2} \to 0 $ as $n\to \infty$,
\begin{align*}
   \hat{\gamma}_\ell(t,t^\prime)=
\frac{1}{n}\sum_{i\in\mathcal{I}_\ell}
\left[X_i(t)-\bar{X}^*_i(t)\right]\left[ X_{i+\ell}(t^\prime)-\bar{X}^*_{i+\ell}(t^\prime)\right],
\end{align*}
with $\mathcal{I}_\ell=\{1,\ldots,n-\ell\}$ if $\ell\geq 0$ and $\mathcal{I}_\ell=\{1-\ell,\ldots,n\}$ if $\ell<0$.
\begin{align*}
   \bar{X}^*_j(t)=\left\{
     \begin{array}{l@{\qquad}l}
      \displaystyle
\frac{1}{\hat{k}_n^*}\sum_{i=1}^{\hat{k}_n^*} X_i(t), & \phantom{\hat{k}^*_n+}1\le  j \le \hat{k}^*_n,  \vspace{.3cm} \\
\displaystyle \frac{1}{n-\hat{k}_n^*} \sum_{i=\hat{k}_n^*+1}^n X_i(t), & \hat{k}^*_n+1 \le  j \le n,
     \end{array}
   \right.
\end{align*}
and $w_\tau$ is a symmetric weight function with bounded support of order $\tau$ satisfying the standard conditions $w_{\tau}(0)=1$, $w_{\tau}(u)=w_{\tau}(-u)$, $w_\tau(u)\le 1$, $w_{\tau}(u)=0$ if $|u|>m$ for some $m>0$, $w_{\tau}$ is continuous, and
\begin{align}\label{order}
0<  \mathfrak{q} =\lim_{x\to 0} x^{-\tau}[1-w_{\tau}(x)] < \infty.
\end{align}
Through $\hat{C}_\varepsilon$, eigenvalue estimates $\hat{\lambda}_1,\ldots,\hat\lambda_n$ of $\lambda_1,\ldots,\lambda_n$ are defined via the integral operator
\begin{align}\label{eigen-emp}
 \hat{\lambda}_\ell \hat{\phi}_\ell(t) &= \int \hat{C}_\varepsilon(t,s)\hat{\phi}_\ell(s)ds.
\end{align}
In order to show consistency of these estimates, a condition supplementary to the weak dependence of the errors $(\varepsilon_i\colon i\in\mathbb{Z})$ given in Assumption \ref{edep} is needed.
\begin{assumption}\label{edep-2}
For some $p>2$, $\ell (\mathbb{E}[\|\varepsilon_i-\varepsilon_{i,\ell}\|^p])^{1/p} \to 0$ as $\ell\to\infty$.
\end{assumption}
Assumption \ref{edep-2} is not necessarily stronger than Assumption \ref{edep}, although both are implied by the simple condition that $(\mathbb{E}[\|\varepsilon_i-\varepsilon_{i,\ell}\|^p])^{1/p}=O(\ell^{-\rho})$ for some $\rho>1$, which is by itself a fairly mild assumption. This condition appears in Horv\'ath et al.\ (2013). The following result holds.
\begin{pro}
Under the conditions of Theorem \ref{th-2} and Assumption \ref{edep-2}, $\hat{C}_\varepsilon$ in \eqref{est-1} is a consistent estimator of $C_\varepsilon$ in $L^2[0,1]^2$. Moreover, for any fixed $d \in \mathbb{N}$, $\max_{1\leq\ell\leq d}|\lambda_\ell - \hat{\lambda}_\ell | =o_P(1)$.
\end{pro}

The verification of this result is given in Lemma A.5 of the online supplement. While the proposition guarantees the large-sample accuracy under a reasonably broad set of conditions, producing the estimate $\hat C_\varepsilon$ and its eigenvalues satisfying \eqref{eigen-emp} in practice requires the choice of a weight function $w_\tau$ and bandwidth $h$. This problem, which is familiar to nonparametric analysis of finite-dimensional time series and spectral density estimation (see, for example, Chapter 7 of Brillinger, 2001), has only recently begun to receive attention in the setting of functional time series.


In the case of long-run covariance function estimation and functional spectral density estimation, H\"ormann and Kokoszka (2010) and Panaretos and Tavakoli (2012) utilized Bartlett and Epanechnikov weight functions (see Bartlett 1946; and Wand and Jones, 1995) with bandwidths of the form $h=n^{1/3}$ and $h=n^{1/5}$, respectively. These choices arise from the well-known fact that taking a bandwidth of the form $h=n^{1/({1+2\tau})}$ maximizes the rate at which the mean-squared normed error of the estimator $\hat{C}_\varepsilon$ tends to zero. The performance of the estimator in finite samples can, however, be affected by strong serial correlation in the data, in which case one should use a larger bandwidth in order to reduce the bias of $\hat{C}_\varepsilon$. An approach that balances these two concerns is to take $h=Mn^{1/({1+2\tau})}$, where the constant $M$ is estimated from the data and increases with the level of serial correlation. It can be shown (see Rice and Shang, 2017) that the optimal constant $M$ in terms of asymptotically minimizing the mean squared normed error of $\hat{C}_\varepsilon$ is of the form
\[
M=\left({2 \tau}\|\mathfrak{q}{C_\varepsilon^{(\tau)}}\|^2\right)^{1/(1+2{ \tau})}\left(\left\{\| C_\varepsilon \|^2 + \left(\int_0^1 C_\varepsilon(u,u)du \right)^2\right\}\int_{-\infty}^\infty w_\tau^2(x)dx   \right)^{-1/(1+2{ \tau})},
\]
where $C_\varepsilon^{(\tau)}$ is related to the $\tau$th derivative of a spectral density operator evaluated at frequency zero. The unknown quantities in $M$ can be estimated using pilot estimates of $C_\varepsilon$ and $C_\varepsilon^{(\tau)}$ to produce an estimated bandwidth $h=\hat{M}n^{1/(1+2\tau)}$. Complete details of this estimation procedure are provided in Section C of the online supplement to the paper.

A comparison of the accuracy in terms of mean-squared normed error of $\hat{C}_\varepsilon$ for a multitude of bandwidth and weight function combinations is provided in Rice and Shang (2017), but a comparative study of how these estimators perform in problems of inference has not been conducted, to the best of our knowledge. With results reported in the online supplement, the proposed break point detection method was compared for all combinations of the Bartlett, Parzen (Parzen, 1957) and a version of the flat-top (Politis and Romano, 1996) weight functions with the four bandwidth choices of $h=n^{1/3}$, $h=n^{1/4}$, $h=n^{1/5}$, and $h=\hat{M}n^{1/(1+2\tau)}$ for the data generating processes considered in the simulation study presented below, as well as some additional processes exhibiting stronger temporal dependence. It was found that when it comes to conducting hypothesis tests and producing confidence intervals as described above with moderately correlated errors, each of these typical choices produced similar results. The difference across weight functions was minuscule, whereas there were some small fluctuations in the empirical sizes of the test of $H_0$ due to the choice of the bandwidth: no more than a 2\% difference when the level was set at 5\% over those FAR processes utilized in Section \ref{sec:sim}, but with expected bigger discrepancies and advantages for the empirical bandwidth when the level of dependence approached non-stationarity. Due to the similarity in performance of each choice for the data generating processes considered below, results are only presented for the bandwidth $h=n^{1/4}$ and the Bartlett weight function below.

\subsection{Computation of critical values}
\label{subsec:imp:critical}

To compute the critical values of the limiting distribution of $T_n$ given in Theorem \ref{inf-main}, say $T$, the following procedure was employed. Based on the estimator $\hat{C}_\varepsilon$, the first $D$ empirical eigenvalues satisfying \eqref{eigen-emp} were computed, where $D$ is taken to be the number of basis elements over which the initial discretely observed functional data are smoothed. By then simulating $D$ independent Brownian bridges on $[0,1]$, $B_\ell(x)$, using the R package {\tt sde},  a realization of $T$ is estimated by
\[
\hat{T} = \sup_{0 \le x \le 1} \sum_{\ell=1}^{D} \hat{\lambda}_\ell B_\ell^2(x).
\]
This estimation is independently repeated $R$ times, and quantiles of the resulting Monte Carlo distribution are used to produce the appropriate cut-offs.  For the results in Sections \ref{sec:sim} and \ref{sec:app}, $R$ was selected to be $1{,}000$.

\subsection{Construction of confidence intervals}
\label{subsec:imp:ci}

This section provides more information on the construction of confidence intervals as defined through Corollary \ref{cor:ci}. To start, let $\hat{\Xi}= \inf\{ x \colon \hat{Q}(x) = \sup_{x^\prime\in\mathbb{R}} \hat{Q}(x^\prime)\}$ be the sample version of $\Xi$, where $\hat Q$ is an estimated version of $Q$ in \eqref{th-2-eq2} obtained by plugging in the natural estimators
\[
\hat{\theta}= \frac{\hat{k}_n^*}{n}
\qquad
\mbox{and}
\qquad
\hat{\sigma}^2 = \intt \hat{C}_\varepsilon(t,t^\prime) \frac{\hat{\delta}_n(t) \hat{\delta}_n(t^\prime)}{\|\hat{\delta}_n\|^2} dtdt^\prime
\]
in place of their respective population counterparts $\theta$ and $\sigma$ as specified in Theorem~\ref{th-2}, where
\[
\hat{\delta}_n = \frac{1}{n-\hat{k}_n^*} \sum_{i=\hat{k}_n^*+1}^n X_i - \frac{1}{\hat{k}_n^*}\sum_{i=1}^{\hat{k}_n^*} X_i,
\]
and $\hat{C}_\varepsilon$ an estimator of $C_\varepsilon$ as discussed in Section \ref{subsec:imp:long-run}. All of these estimators are consistent under the conditions of Theorem \ref{th-2}; see Lemma A.5 of the online supplement. Let $\hat{\Xi}_q$ denote the $q$th quantile of the distribution of $\hat{\Xi}$.

\begin{theorem}\label{ci-th}
Under the conditions of Theorem \ref{th-2} and Assumption \ref{edep-2}, for $\alpha\in(0,1)$, the random interval
\[
\left( \hat{k}^*_n - \frac{\hat{\Xi}_{1-\alpha/2}}{\|\hat{\delta}_n\|^{2}},\hat{k}^*_n - \frac{\hat{\Xi}_{\alpha/2}}{\|\hat{\delta}_n\|^{2}}\right)
\]
is an asymptotic $1-\alpha$ confidence interval for $k^*$.
\end{theorem}

Note that the construction of confidence intervals is aided by the use of the exact form of the maximizers in the limit of Theorem \ref{th-2} as derived in Bhattacharya and Brockwell (1976) and Stryhn (1996), see the supplemental material for more. Since $\sigma^2$ and $\hat{\sigma}^2$ are respectively bounded from above by $\lambda_1$ and $\hat{\lambda}_1$, the largest eigenvalues of the integral operators with kernels $C_\varepsilon$ and $\hat{C}_\varepsilon$, a conservative confidence interval is obtained by replacing $\hat{\sigma}^2$ with $\hat{\lambda}_1$.

\section{Simulation Study}
\label{sec:sim}

\subsection{Setting}
\label{sec:sim:setting}

Following the construction of the data generating processes (DGP's) in Aue et al.\ (2015), $n$ functional data objects were generated using $D=21$ Fourier basis functions $v_1,\ldots,v_D$ on the unit interval $[0,1]$. The choice of $D$ corresponds to our study of yearly Australian temperature curves constructed from daily minimum temperature observations that were initially smoothed over this basis. Qualitatively these results remain valid for larger values of $D$. Without loss of generality, the initial mean curve $\mu$ in \ref{model-1} is assumed to be the zero function. Independent curves were then generated according to
\[
\zeta_i=\sum_{\ell=1}^DN_{i,\ell}v_\ell,
\]
where the $N_{i,\ell}$ are independent normal random variables with standard deviations $\bm{\sigma}=(\sigma_\ell\colon \ell=1,\ldots,D)$ used to mimic various decays for the eigenvalues of the covariance and long-run covariance operators. Three distinct situations were considered:
\begin{itemize}\itemsep-.4ex
\item {\it Setting 1:}\/ the errors are finite dimensional, using $\sigma_\ell=1$ for $\ell=1,2,3$ and $\sigma_\ell=0$ for $\ell=3,\ldots,D$;
\item {\it Setting 2:}\/ mimics a fast decay of eigenvalues, using $\bm{\sigma}=(3^{-\ell}\colon \ell=1,\ldots,D)$;
\item {\it Setting 3:}\/ mimics a slow decay of eigenvalues, using $\bm{\sigma}=(\ell^{-1}\colon \ell=1,\ldots,D)$.
\end{itemize}
Note that the last setting, inspired by the data analysis reported in Section \ref{sec:app}, is not to be taken asymptotically in $D$. Rather it is meant to model a slow decay of the finitely many initial eigenvalues without intending to prescribe the behavior for $D$ tending to infinity.

As innovations, independent curves $\varepsilon_i=\zeta_i$, $i=1,\ldots,n$, were used. To explore the effect of temporal dependence on the break point estimators, functional autoregressive curves were also considered, which are widely used to model serial correlation of functional data, see Besse et al.\ (2000) and Antoniadis and Sapatinas (2003). First-order functional autoregressions $\varepsilon_i=\Psi\varepsilon_{i-1}+\zeta_i$, $i=1,\ldots,n$, were generated (using a burn-in period of 100 initial curves that were discarded). The operator was set up as $\Psi=\kappa\Psi_0$, where the random operator $\Psi_0$ is represented by a $D\times D$ matrix whose entries consist of independent, centered normal random variables with standard deviations given by $\bm{\sigma}\bm{\sigma}^\prime$ as specified by Settings 1--3. A scaling was applied to achieve $\|\Psi_0\|=1$. The constant $\kappa$ can then be used to adjust the strength of the temporal dependence. To ensure stationarity of the time series, $|\kappa|=0.5$ was selected.

To highlight the effect of the distribution of the break function across eigendirections as well as its size relative to the noise level, the following arrangements were made. A class of break functions was studied given by
\begin{equation}
\label{eq:break_fct}
\delta_m=\delta_{m,c}=\sqrt{c}\delta_m^*, \qquad \delta_m^*=\frac{1}{\sqrt{m}}\sum_{\ell=1}^mv_\ell,
\qquad m=1,\ldots,D,
\end{equation}
where the normalization ensures that all $\delta_m^*$ have unit norm. The role of $c$ is explained below. Note that $\delta_1$ represents the case of a break only in the leading eigendirection of the errors. On the other end of the spectrum is $\delta_D$ describing the case of a break that affects all eigendirections uniformly. To relate break size to the natural fluctuations in the innovations, the signal-to-noise ratio
\[
\mathrm{SNR}
=\frac{\theta(1-\theta)\|\delta_{m}\|^2}
{\mathrm{tr}(C_\varepsilon)}
=c\frac{\theta(1-\theta)}
{\mathrm{tr}(C_\varepsilon)},
\]
was used, where $\theta$ denotes the relative location of the break date and $C_\varepsilon$ the long-run covariance operator of the $\varepsilon_i$. (Note that since $\|\delta^*_m\|=1$, in the adopted formulation $\mathrm{SNR}$ does not depend on $m$.) Results are reported choosing $c$ to maintain a prescribed SNR.

Finally, in order to mitigate the effect of the particular shape of the Fourier basis functions and the ordering $v_1,\ldots,v_D$ on the performance of the various procedures, a random permutation $\pi$ was applied to $1,\ldots,D$ before each simulation run, and the experiment was performed as described above using the permuted ordering $v_{\pi(1)},\ldots,v_{\pi(D)}$. Combining the previous paragraphs, functional curves $y_i=\delta_m\mathds{1}\{i > k^*\} + \varepsilon_i$, $i=1,\ldots,n$, according to \eqref{model-1} were generated for $k^*=\lfloor\theta n\rfloor$ with $\theta=0$ (null hypothesis), and $\theta=0.25$ and $0.5$ (alternative). Both the fully functional procedure and its fPCA counterparts were applied to a variety of settings, with outcomes reported in subsequent sections. 
All results  are based on 1000 runs of the simulation experiments. Additional complementary simulation evidence is presented in the online supplement.

\subsection{Level and power of the detection procedures}
\label{sec:sim:level+power}

In this section, the level and power of the proposed detection procedure are compared to the two fPCA-based methods introduced in Section \ref{sec:fpca_methods}. In particular, the fPCA-based detector \eqref{eq:fPCA_detector} was run with three levels of total variation explained (TVE), namely 85\%, 90\% and 95\%. The change-aligned detection procedure \eqref{eq:aligned} was set up as in Torgovitski (2016). Critical values for the proposed fully functional detection procedure were obtained through simulation from the limit distribution under the null hypothesis as provided in Theorem~\ref{inf-main}.

Table \ref{tab:level} provides the levels for the various detection procedures for the three settings of eigenvalue decays, and iid and FAR(1) data generating processes. For the FAR(1) case, the long-run covariance operator was estimated following the recommendations given in Rice and Shang (2017). The sample sizes under consideration were $n=50$ and $n=100$. It can be seen that, even for these rather small to moderate sample sizes, the proposed method kept levels reasonably well across all specifications. This is true to a lesser extent also for the fPCA-based procedures, while the change-aligned version produced the most variable results. The fPCA-based procedures depend, by construction, more explicitly on the behavior of the eigenvalues with levels well adjusted in case of a fast decay. The proposed procedure is fairly robust in all settings.

\begin{table}[h]
\centering
\begin{tabular}{c@{\qquad}c@{\qquad}c@{\qquad}c@{\qquad}c@{\qquad}c@{\qquad}c@{\qquad}c}
\hline
Setting &DGP &$n$ & Proposed & TVE 85\% & TVE 90\% & TVE 95\% & Aligned  \\ \hline
1 & iid & \phantom{1}50 & 0.08 & 0.02 & 0.03 & 0.03 & 0.02 \\
&&100 &  0.06 & 0.07 & 0.06 & 0.06 & 0.05 \\[.1cm]
& FAR(1) & \phantom{1}50 & 0.07 & 0.03 & 0.03 & 0.02 & 0.00 \\
&&100 & 0.05 & 0.07 & 0.06 & 0.07 & 0.02 \\
\hline
2 & iid & \phantom{1}50 & 0.07 & 0.06 & 0.06 & 0.06 & 0.09 \\
&&100 & 0.06 & 0.05 & 0.06 & 0.07 & 0.05 \\[.1cm]
& FAR(1) & \phantom{1}50 & 0.07 & 0.04 & 0.04 & 0.05 & 0.11 \\
&&100 & 0.06 & 0.05 & 0.05 & 0.05 & 0.05 \\
\hline
3 & iid & \phantom{1}50 & 0.04 & 0.02 & 0.02 & 0.00 & 0.00 \\
&&100 & 0.05 & 0.07 & 0.03 & 0.02 & 0.01 \\[.1cm]
& FAR(1) & \phantom{1}50 & 0.03 & 0.01 & 0.03 & 0.00 & 0.00 \\
&&100 & 0.05 & 0.02 & 0.02 & 0.02 & 0.00 \\
\hline
\end{tabular}
\caption{Empirical sizes for the various detection procedures for two data generation processes. The nominal level was $\alpha=0.05$. }
\label{tab:level}
\end{table}

\begin{figure}[h!]
\vspace{.4cm}
\begin{center}
\includegraphics[width=\textwidth]{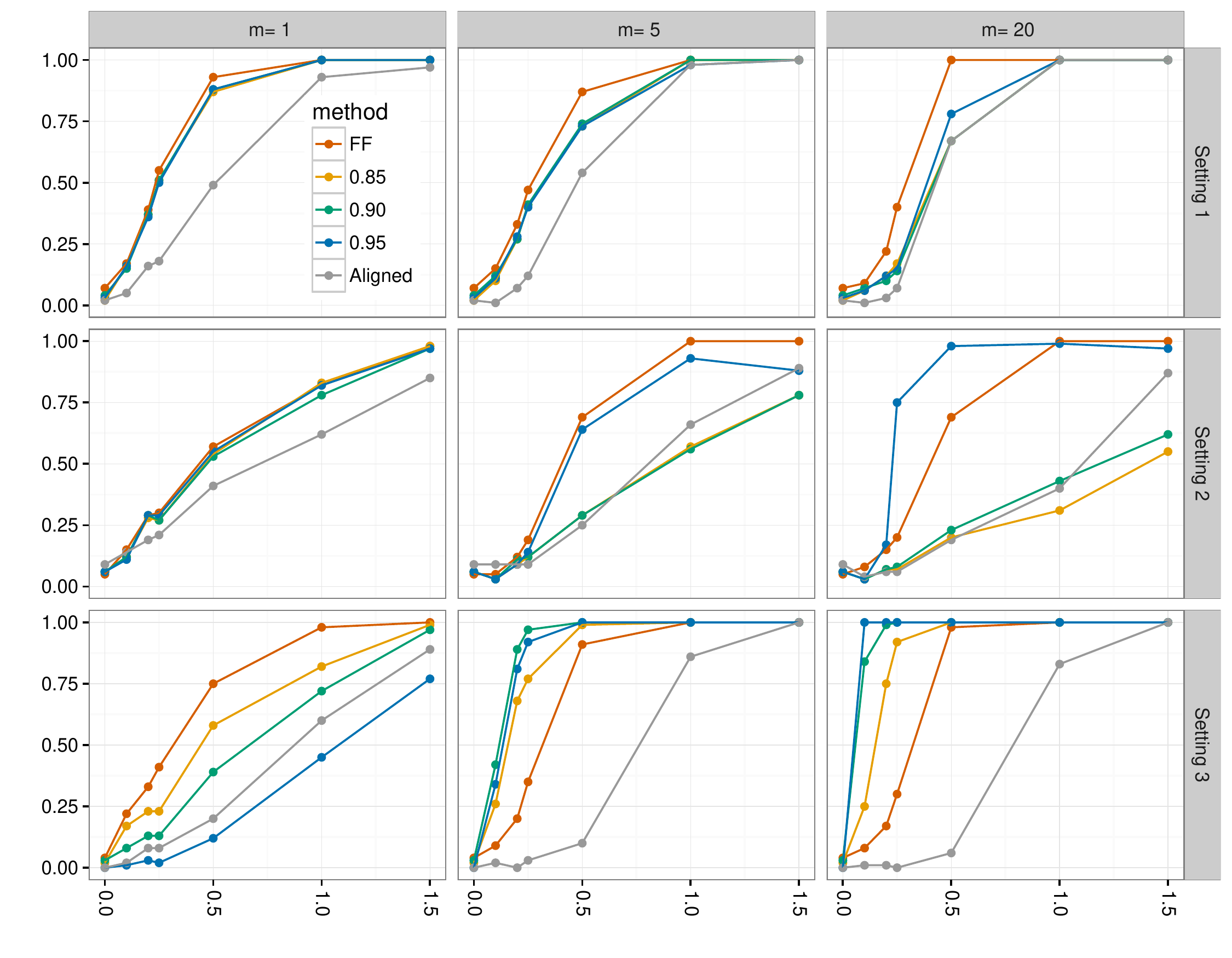}
\end{center}
\vspace{-0.8cm}
\caption{Power curves for the various break detection procedures for three different forms of the break functions indexed my $m$ and the three eigenvalues settings for $n=50$ and independent errors. The $x$-axis gives different choices of SNR. Observe that ``FF'' refers to the proposed fully functional method, ``0.85'', ``0.90`` and ``0.95`` correspond to the three levels of TVE in the fPCA procedures, and ``Aligned'' to the method of Torgovitski (2016).
\label{fig:power-iid}}
\end{figure}

To examine the power of the detection procedures in finite samples, the break functions $\delta_m$ in \eqref{eq:break_fct} were inserted as described in Section \ref{sec:sim:setting} with scalings $c$ so that the SNR varied between 0, 0.1, 0.2, 0.3, 0.5, 1 and 1.5. The empirical rejection rates out of 1000 simulations for each test statistic described above are reported as power curves in Figure \ref{fig:power-iid} when the errors in \eqref{model-1} are iid and distributed according to each of Settings 1, 2, and 3. The sample size in the figure is $n=50$ and the number of components $m$ in the break functions $\delta_m$ are 1, 5, and 20. Note that the plots in Figure \ref{fig:power-iid} are not size corrected because it would not qualitatively change the outcomes. Further simulation evidence is provided in the online supplement. The findings of these simulations can be summarized as follows: \vspace{-.2cm}
\begin{itemize}\itemsep-.2ex
  \item The change aligned test of Torgovitski (2016) was usually outperformed by both the fPCA and fully functional methods for most of the DGP's and sample sizes under consideration.
  \item The power for the fully functional detection procedure was observed to improve as $m$ increased, as predicted by the theory. Moreover, when the change was largely orthogonal to the errors, as in Setting 1 with $m=20$, the expected advantage of the fully functional method over the dimension reduction based approaches materialized.
  \item A particularly interesting example to examine is when $m=1$ under Setting 3 (with slowly decaying eigenvalues). One notices in this case that, although the change lied fully in the direction of the leading principal component of the errors, the fPCA-based methods were outperformed by the fully functional method, and additionally their performance decayed as TVE increased. Here, the slow decay of eigenvalues adversely affects the fPCA procedure. This contrasts, for example, with the case when $m=20$ under Setting 2 (with fast decaying eigenvalues), when the fPCA method improved as TVE increased, and ultimately outperformed the fully functional method. This demonstrates that the fPCA method is not guaranteed to beat the proposed detection procedure even when the break is  in the leading eigendirection. Moreover, this particular case highlights the fact that increasing TVE may not always lead to improved performance. Note also that this example seems to match well with the situation encountered in an application to Australian temperature curves presented in Section~\ref{sec:app}.
  \item In additional simulations reported in the online supplement, the expected improvement in power when $n$ increased was noticed. Additionally, no more power loss than is typical was observed when the model errors are serially correlated rather than independent and identically distributed.
\end{itemize}

\subsection{Performance of the break dating procedures}
\label{sec:sim:dating}

\begin{figure}[h!]
\vspace{.4cm}
\begin{center}
\includegraphics[width=\textwidth]{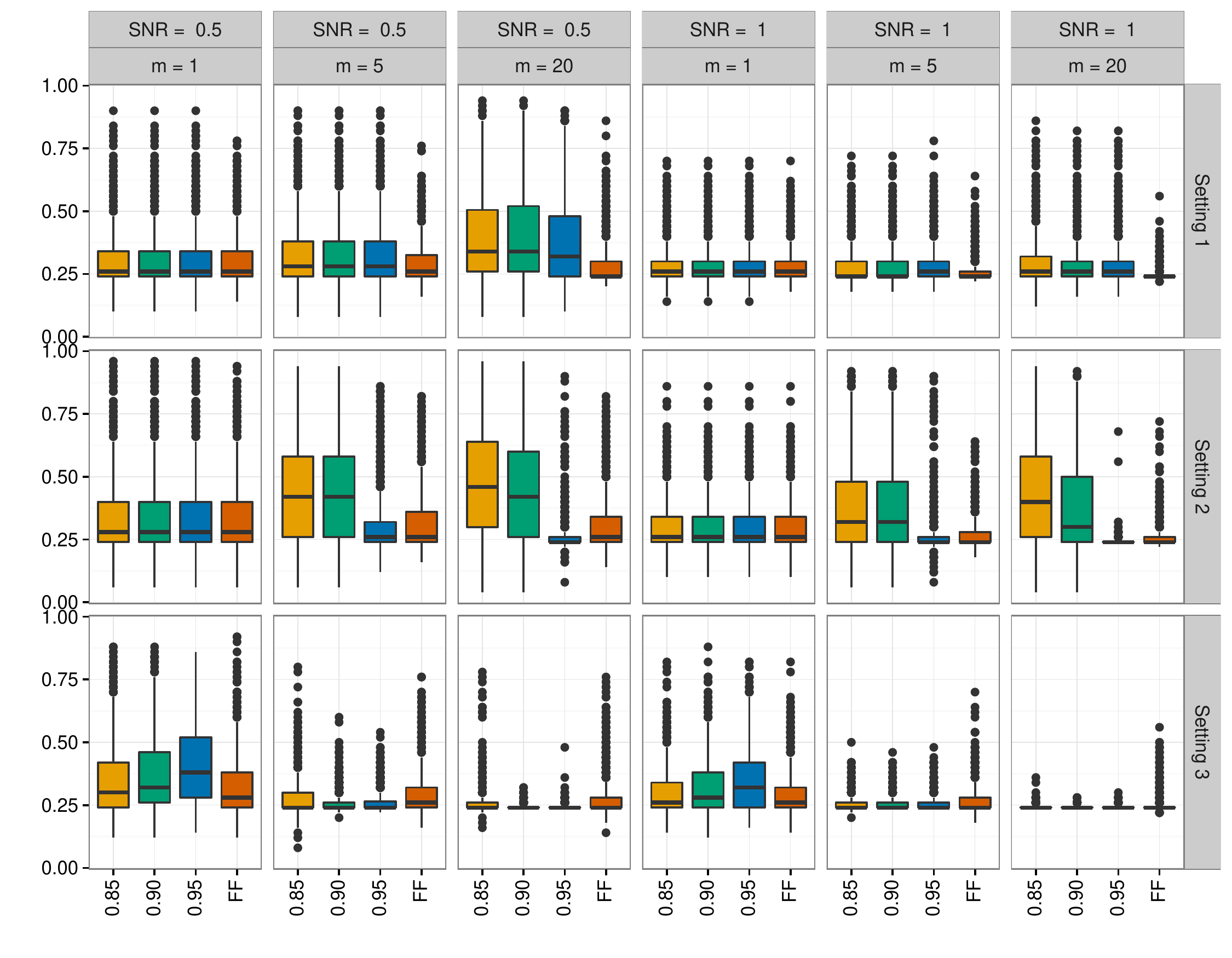}
\end{center}
\vspace{-0.8cm}
\caption{Boxplots for the various break dating procedures for three different forms of the break functions indexed my $m$ and two choices of SNR for the three eigenvalues settings, sample size $n=50$ and independent errors. Labeling of the procedures is as in Figure \ref{fig:power-iid}.
\label{fig:estimation-iid}}
\end{figure}

In order to study the empirical properties of the break date estimator $\hat{k}^*_n$, the break functions $\delta_m$ specified in \eqref{eq:break_fct} of Section \ref{sec:sim:setting} were utilized again with scaling $c$ chosen to yield SNR values of 0.5 and 1. The break date was inserted at $\theta=0.25$, so that the samples before and after the break have a ratio of 1 to 3. As in the previous section, focus is on the small sample size $n=50$. The results from additional settings are reported in the online supplement. For each setting and choice of $m$, the estimators $\hat k_n^*$ and $\tilde k_n^*$ for $k^*$ were computed for the proposed and the fPCA methods, respectively in 1000 independent simulation runs. The results are summarized in the form of box plots in Figure \ref{fig:estimation-iid}.

Overall, the proposed method is observed to be competitive, with box plots being narrower or of the same width as those constructed from the fPCA counterparts. It can be seen that the accuracy of the fully functional break date procedure improved for increasing $m$, spreading the break across a larger number of directions. As expected, the performance of the fPCA procedure was sensitive to the choice of TVE, in a way that often only the best selection of TVE was competitive with the fully functional method. Moreover, in analogy to the same phenomenon observed in the power study, the fully functional procedure enjoys an advantage when the break loads entirely on the first eigenfunction ($m=1$) for slowly decaying eigenvalues of the covariance operator (Setting 3).

\begin{figure}[h!]
\vspace{.4cm}
\begin{center}
\includegraphics[width=.75\textwidth]{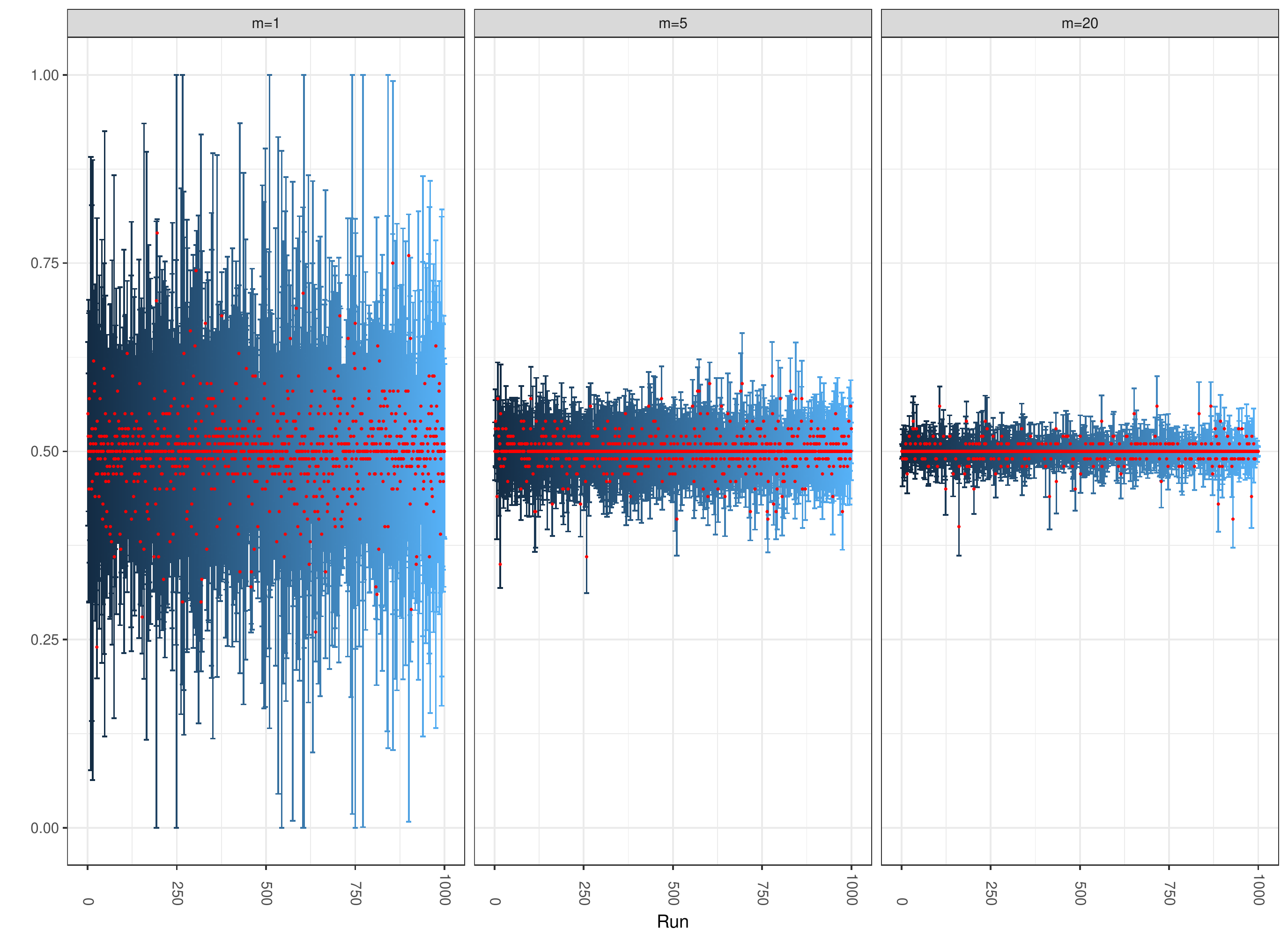} 
\end{center}
\vspace{-0.5cm}
\caption{Confidence intervals constructed from the fully functional break dating procedure across 1000 simulation runs for Setting 3, sample size $n=100$, and three types of break functions indexed by $m$ with SNR set to $0.5$. For each run, the blue line gives the 95\% confidence interval and the red dot the estimated break date.
\label{fig:ci-iid}}
\end{figure}

The confidence intervals computed from Theorem \ref{ci-th} are seen to be conservative. As already pointed out after Corollary \ref{ci-f}, this is due to the fact, that they are based on an asymptotic analysis assuming a shrinking break. For illustration purposes, since this will prove relevant in Section \ref{sec:app}, Figure \ref{fig:ci-iid} gives 95\% confidence intervals for the case of Setting 3 with independent errors and sample size $n=100$. The break function $\delta_m$ is inserted in the middle ($\theta=0.5$), using $m=1$, $5$ and $20$ as before. The plots provide further evidence for the theory, as the confidence intervals get significantly narrower when the break function is distributed across a larger number of directions. The case $m=1$ leads to the widest confidence intervals, which for this case are of little practical relevance. Larger sample sizes and higher SNR lead to the expected improvements, but are not shown here to conserve space. To improve the width of the confidence intervals for small sample sizes and/or small SNR's, one might entertain some jackknife or bootstrap modifications. This might be pursued in detail elsewhere.

\subsection{Heavy tails}
\label{sec:sim:heavy}

The heavy tail case is only considered for independent curves in Settings 2 with fast decay of eigenvalues of the innovations and break function specified by $\delta_m$ in \eqref{eq:break_fct} with $m=1$, $5$ and $20$ as before. Settings 1 and 3 produce results more in favor of the proposed method. Instead of the normal distributions specified in Section \ref{sec:sim:setting}, $\zeta_1,\ldots,\zeta_n$ were chosen to be $t$-distributed with 2, 3 and 4 degrees of freedom and $\varepsilon_1,\ldots,\varepsilon_{100}$ were defined accordingly. Modifications of the simulation settings presented in this section could potentially be useful for applications to intra-day financial data such as the Microsoft intra-day return data presented as part of the online supplement. Due to the reduced number of finite moments in this setting, the fPCA-based procedure is not theoretically justified, while the fully functional procedure is not justified only for the case of two degrees of freedom.

\begin{figure}[h!]
\vspace{.3cm}
\begin{center}
\includegraphics[width=.8\textwidth]{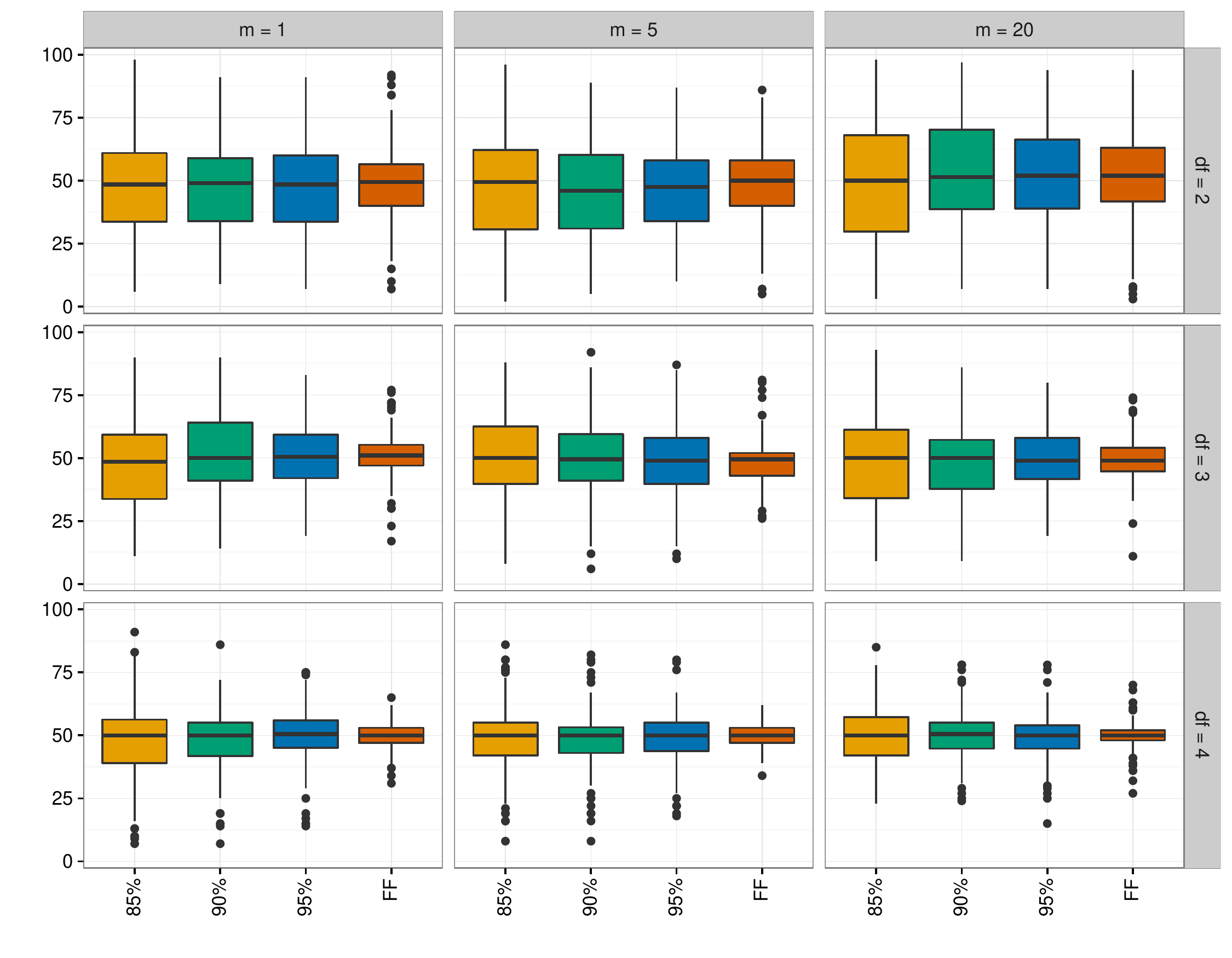}
\end{center}
\vspace{-0.8cm}%
\caption{Boxplots for the various break dating procedures for three different forms of the break functions indexed my $m$ and $t$-distributed innovations with 2, 3 and 4 degrees of freedom for Setting 2, sample size $n=100$, $k^*=50$ and independent errors. Labeling of the procedures is as in Figure \ref{fig:power-iid}.
\label{fig:df}}
\end{figure}

Results in Figure \ref{fig:df} are given for $n=100$, $k^*=50$. The summary statistics show the proposed method to be superior in all cases. The proposed method looks in general more favorable in the heavy-tail case than in the time series case of the previous section due to the deteriorated performance in estimating eigenvalues and eigenfunctions.
It can be seen that in all cases the fPCA-based procedure fails to produce reasonable results. The performance is worst for $\mathrm{df}=2$ and somewhat comparable for $\mathrm{df}=3$ and $\mathrm{df}=4$. The proposed method is seen to work for the latter two cases but its performance deteriorates somewhat for $\mathrm{df}=2$, a situation that is not theoretically justified.


\section{Application to annual temperature curves}
\label{sec:app}

In this section, the proposed methodology is applied to annual temperature curves from eight measuring stations in Australia. More precisely, the raw data consists of 365 (366) daily measurements of minimum temperatures that were converted into functional objects using 21 Fourier basis functions. The observations for each of the eight stations are recorded over different time spans, roughly equaling 100 years. The data may be downloaded from The Australian Bureau of Meteorology at the URL \url{www.bom.gov.au}. For each case, the fully functional break detection procedure rejected the null hypothesis of no structural break in the mean function. Consequently, both functional break dating procedure and fPCA counterpart were applied to locate the time of the mean break. Information on all stations under consideration is provided in Table \ref{tab:app:stations}. More details may be found in the online supplement.


\begin{table}[h!]
\vspace{.6cm}
\begin{center}
\begin{tabular}{l@{\qquad}c@{\qquad}c@{\qquad}c@{\qquad}c}
\hline
Station\phantom{$\hat{\hat\hat{k^*}}k_{n_n}$} & Range & $\hat k_n^*$ (year) & CI (years)& Range of $\tilde k^*_n$ (year) \\
\hline
Sydney (Observatory Hill) & 1959--2012 & 1991 & (1981, 1994) & 1983, {\bf 1991} \\
Melbourne (Regional Office) & 1855--2012 & 1998 & (1989, 2000) & 1996, {\bf 1998} \\
Boulia Airport & 1888--2012 & 1978 & (1954,\,1981) & {\bf 1978}\\
Cape Otway Lighthouse & 1864--2012 & 1999 & (1949, 2005) & 1999, {\bf 2000} \\
Gayndah Post Office & 1893--2009 & 1962 & (1952, 1966) & 1953, 1962, {\bf 1968} \\
Gunnedah Pool & 1876--2011& 1985 & (1935, 1992) & 1979, 1984, {\bf 1985}, 1986 \\
Hobart (Ellerslie Road) & 1882--2011 & 1966 & (1957, 1969) & {\bf 1966}, 1967, 1968, 1969\\
Robe Comparison & 1884--2011 & 1981 & (1954, 1985) & 1969, 1974, {\bf 1981}
\\ \hline
\end{tabular}
\caption{\label{tab:app:stations} Summary of results for eight Australian measuring stations. The column labeled $\hat k^*_n$ reports the estimated break date using the fully functional method, CI gives the corresponding 95\% confidence interval. This is contrasted with the range of break date estimates obtained from using fPCA methods with dimension of the projection space $d=1,\ldots,10$. The year in bold is the most frequently chosen break date.
}
\end{center}
\end{table}

\begin{figure}[h!]
\vspace{-.4cm}
\begin{center}
\includegraphics[width=.41\textwidth]{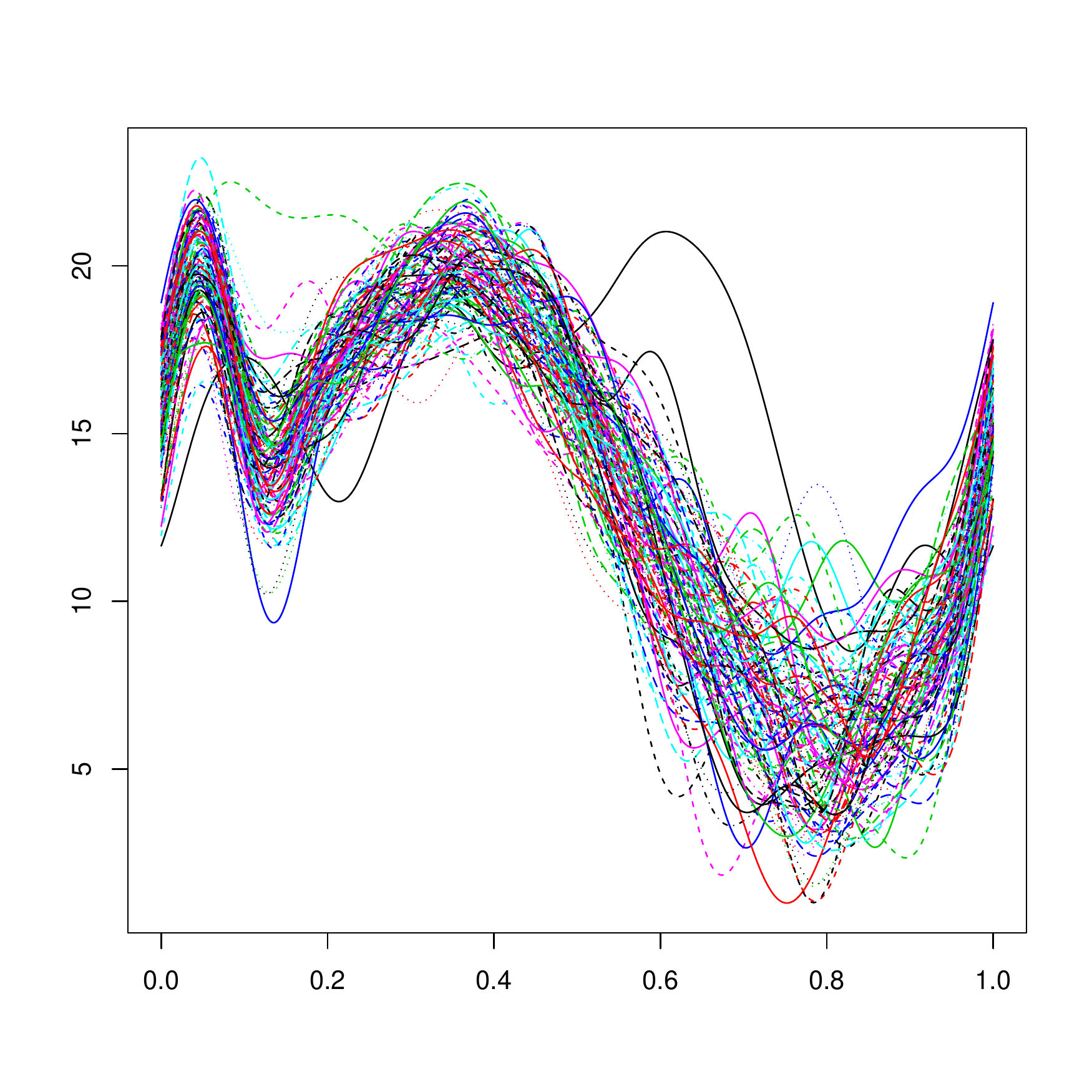}
\includegraphics[width=.41\textwidth]{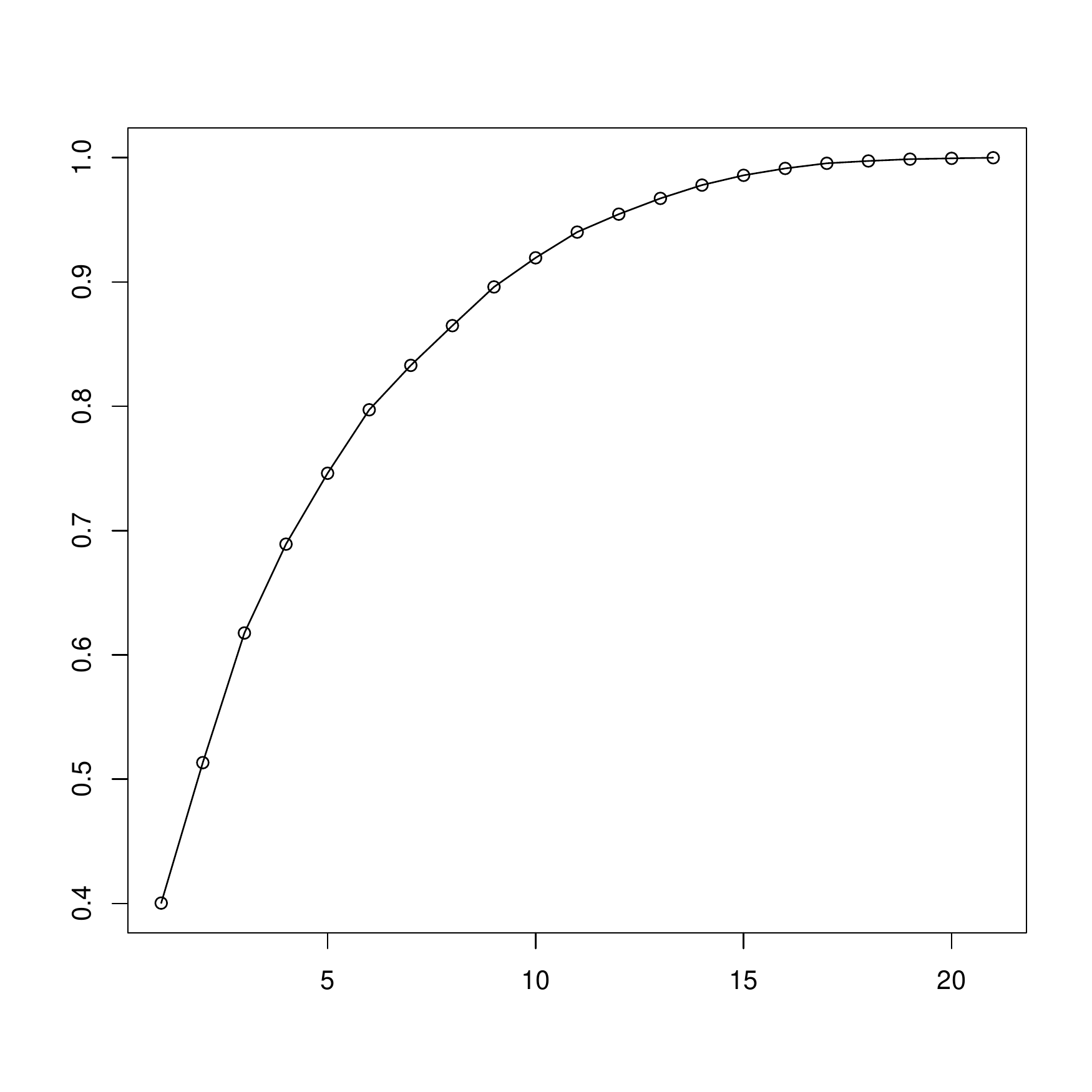} \\[-.3cm]
\includegraphics[width=.41\textwidth]{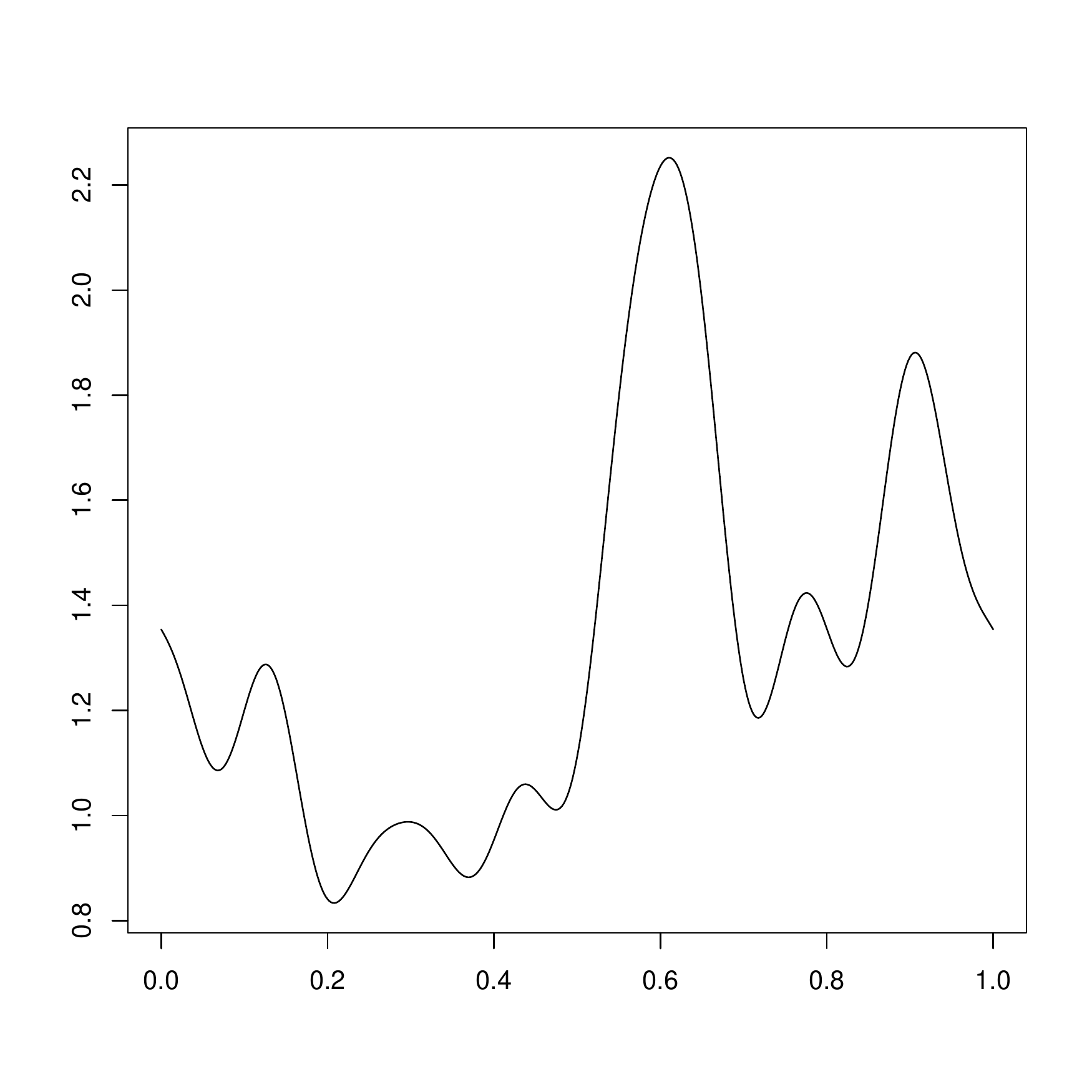}
\includegraphics[width=.41\textwidth]{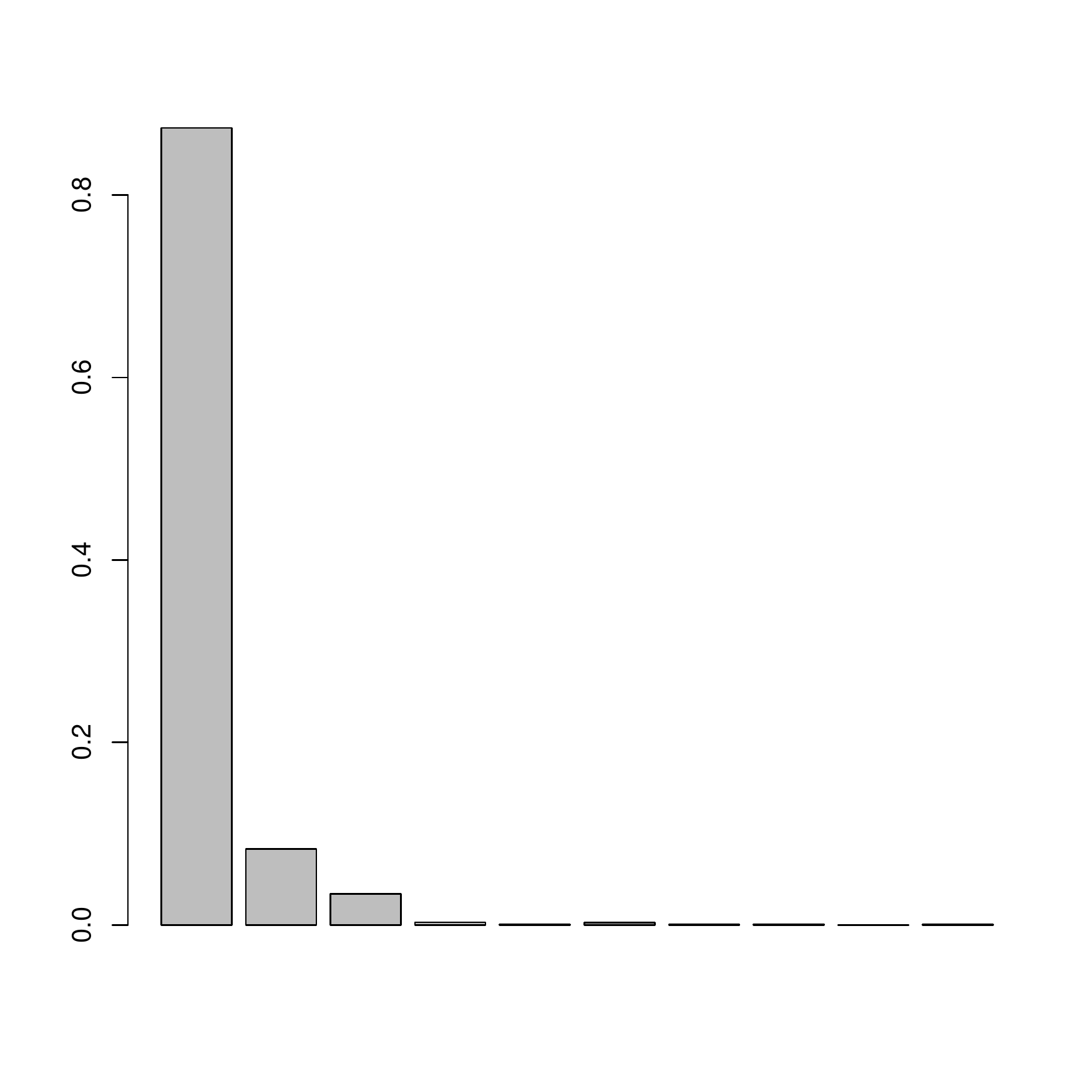} \\[-.3cm]
\includegraphics[width=.41\textwidth]{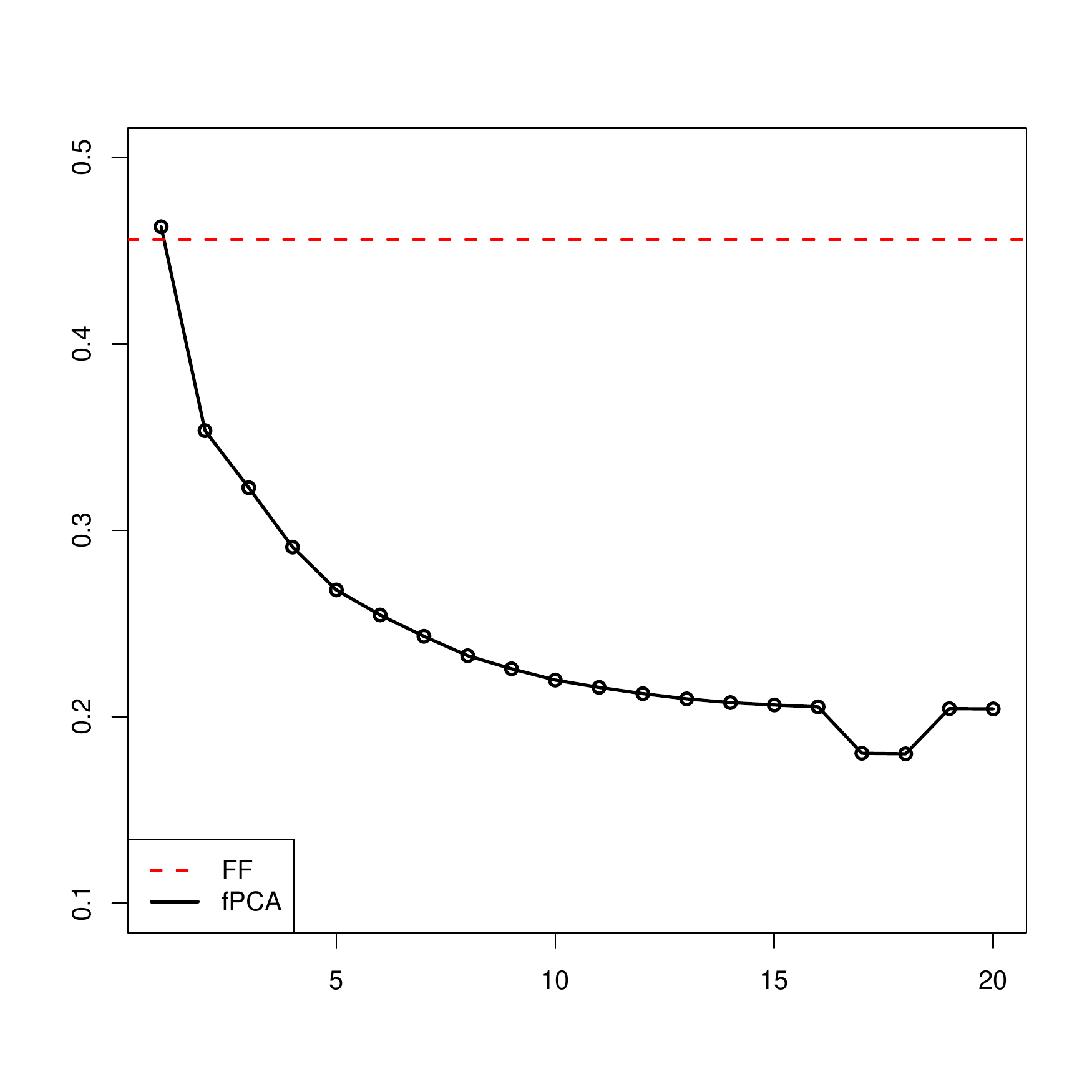}
\end{center}
\vspace{-0.8cm}
\caption{Upper panel: Time series plot of annual temperature profiles at Gayndah Post Office (left) and scree plot of eigenvalues from the sample covariance operator of the Gayndah Post Office temperature profiles (right). Middle panel: Estimated break function $\hat\delta$ (left) and proportion of variation in $\|\hat\delta\|$ explained by the $\ell$th sample eigenfunction (right). Lower panel: Estimated SNR for the fully functional procedure (straight line) and for the fPCA procedure across varying $d$.
\label{fig:app:tsplot}}
\end{figure}

In the following the station Gayndah Post Office is singled out and discussed in more detail. The time series plot of $n=116$ annual curves recorded in degree Celsius at this station from 1893 to 2009 are given in the upper left panel of Figure \ref{fig:app:tsplot}. They exhibit the temperature profile typical for Australia, with higher temperatures in the beginning and end of the year. The corresponding scree plot of sample eigenvalues in the upper right panel of the same figure indicates a slow decay, which Setting 3 in Section \ref{sec:sim} sought to mimic. The $p$-value of the fully functional detection procedure for this station was $0.008$. Table \ref{tab:app:stations} reports the break date estimate for the fully functional procedure as $1962$ and gives a 95\% confidence interval spanning the years from 1952 to 1966. In the range considered, the fPCA procedure chose three different years as break dates, namely 1953 (corresponding to $d=1$ and $\mathrm{TVE}=0.40$), 1962 (for $d=3$ and $\mathrm{TVE}=0.62$), and 1968 (for all other choices of $d$ with $\mathrm{TVE}$ reaching 0.92 at $d=10$). It can therefore be seen that, for any reasonable choice of $\mathrm{TVE}$, the fPCA break date estimate leads to a year that is not included in the 95\% confidence interval obtained from the fully functional methodology, even those were shown to be conservative in Section \ref{sec:sim}. The estimated break function is displayed in the middle panel of Figure \ref{fig:app:tsplot}. Almost 90\% of the variation in $\|\hat\delta\|$ is explained by the first sample eigenfunction, with a rapid decay of contributions from higher sample eigenfunctions. This is displayed in the middle panel of Figure \ref{fig:app:tsplot}. The situation is therefore indeed similar to the case displayed in the lower left panels of Figures \ref{fig:power-iid} and \ref{fig:estimation-iid}, which corresponds to slow decay of eigenfunctions and a break occurring predominantly in the direction of the first mode of variation. That this is a situation beneficial to the proposed procedure is further highlighted in the lower panel of Figure \ref{fig:app:tsplot}. Here it can be seen that the estimated SNR of the sample break function decreases significantly with the inclusion of further sample eigenvalues and eigenfunctions into the analysis. In particular, the estimated SNR's are, for $d>1$, noticeably smaller than the estimated SNR obtained from the fully functional procedure.

The application shows that, while both fully functional and fPCA procedures often work similarly in practice, there are cases when they differ substantially. In the situation discussed in this section, there is evidence to believe that the fully functional method is perhaps more trustworthy. The results of the data application used in combination with the simulation analysis show that one can do worse than the proposed procedure but not obviously better.

\section{Conclusions}
\label{sec:conclusions}

In this paper, a fully functional methodology was introduced to detect and date mean curve breaks for functional data. The assumptions made allow for time series specifications of the curves and are formulated using the optimal rates for approximations of the data with $\ell$-dependent sequences. The assumptions are notably weaker than those usually made in the fPCA context and include heavy-tailed functional observations, making the asymptotic theory developed here widely applicable. In a comprehensive simulation study it is shown that the fully functional method tends to perform better than its fPCA counterpart, with significant performance gains for breaks that do not align well with the directions specified by the largest (few) eigenvalue(s) of the data covariance operator, but also in a number of subtler situations such as breaks concentrated on the first eigendirection with slowly decaying eigenvalues. It is shown in an application to annual temperature curves that the latter situation can be of practical relevance. 
More generally, this work provides an in-depth study in a specific context of the overarching principle that whenever the signal of interest is not dominant or is ``sparse", in the sense that it is not entirely contained in the leading principal components, then alternatives to dimension reduction based methods should be considered and are likely more effective.

\end{document}